\documentclass[12pt,a4paper]{article}
\usepackage[usenames]{color}
\usepackage{amsfonts, amssymb, amsmath, amsthm, latexsym,amscd, xypic}
\usepackage[dvips,matrix,ps,color,line,graph]{xy}

\begin{document}


\title{Fibered Manifolds, Natural Bundles, Structured Sets, G-Sets and all that: 
The Hole Story from Space Time to Elementary Particles}
\author{J. Stachel\thanks{Department of Physics and Center for Einstein Studies, 
Boston University} \hspace{2pt}, M. Iftime\thanks{Department of Mathematics, School of Arts \& Sciences, MCPHS Boston}}
\date{}
\maketitle

\begin{flushright}
{\scriptsize [M]ath\'{e}maticiens et physiciens ont pris conscience, depuis longtemps 
d\'{e}j\`{a}, du fait\\ 
que les  espaces fibr\'{e}s constituent un cadre de pens\'{e}e fondamental 
pour la relativit\'{e}, \\
comme ils le font d'ailleurs aussi pour la m\'{e}canique analytique classique. \\({\it Lichnerowicz, Espaces Fibres et Espace-Temps, GRG, vol {\bf 1}. No. {\bf 3}, pp 235-245)}}
\end{flushright}

\begin{abstract}

In this paper we review the hole argument for the space-time points and elementary particles  and generalize the hole argument to include all geometric object fields and diffeomorphisms; and, by application of forgetful functors to abstract from differentiability and even continuity, the hole argument is applied to a much wider class of mathematical objects.
We discuss the problem concerning the individuation of the objects in more general settings
such that fibered manifolds, fibered sets and n-ary relations.

\end{abstract}
\pagebreak

\tableofcontents
\pagebreak
\section{Introduction}


We begin with a review of the original hole argument, which deals with the 
space-time of general relativity, i.e., a four-dimensional manifold $M$ together with 
a pseudo-metrical field $g_{\mu \nu }$ with Minkowski signature (from now on we omit the prefix 
``pseudo''). \footnote{For the history of the hole argument, see \cite{Stachel1989}, 
for full accounts of the hole argument,see \cite{Stachel1993} and the references to earlier literature therein.}
This field represents not only the chrono-geometrical structure of space-time, but also the
potentials for the inertio-gravitational field.\footnote{See e.g., \cite{Stachel1994}} 
There may also be any number of other tensor 
fields on the manifold, representing non-gravitational fields and/or matter and acting as sources 
of the metrical field in the inhomogeneous Einstein equations
\footnote{In most cases the metric field appears on both left and right sides of Einstein equations, so the space time structure and the source fields in space-time 
constitute a dynamical system, the equations of which can only be solved together. For some exceptions (\cite{Stachel1969})}
Let the manifold $M$ contain a 
{\em hole}, $H$, i.e., an open region, on which the metric field is the only one present, so that  
inside $H$ the metric obeys the homogeneous Einstein equations. 
The hole argument involves {\em diffeomorphisms} of $M$
\footnote{That is the 1:1 bi-continuous and differentiable mappings of the manifold 
onto itself, i.e., its differentiable automorphisms. We assume the order of differentiability needed for the equations in questions \cite{Crampin}, pp. 240}  
and the mappings they induce on the metric and any other tensor fields.
\footnote{The Appendix shows in detail how to construct these 
induced mappings for any geometric object field.}

The Einstein field equations are covariant. By definition, a set of field equations is said 
to be {\em covariant} if, whenever a set of fields is a solution of these equations, then any other 
set obtained from the first by the mapping induced by {\em any} diffeomorphism of $M$ also 
satisfies these equations. We refer to such fields as dragged-(carried-) along fields. 
Suppose the points of $M$ are individuated independently of the fields in question
and the field equations are covariant then
even if all the fields in question are specified everywhere outside of and on the boundary of the hole $H$ and even if all the normal derivatives of these fields up to any finite order are also specified  on the boundary, then the metric field inside $H$ is still not determined uniquely
no matter how small the hole $H$. 
  
The proof is simply to note that, given any solution inside the hole, an 
unlimited number of other solutions can be generated from it by those diffeomorphisms that 
{\em reduce to the identity} on $M-H$ (and any number of derivatives of which also reduce to the 
identity on the boundary), but {\em differ from the identity} inside the hole $H$.

In this proof, the distinctness of the diffeomorphically-related solutions depends crucially on the assumption that {\em the points of the manifold inside 
the hole are individuated independently of the fields inside $H$ }
(i.e., the metric field in the original hole argument). 
Then each of the diffeomorphically-related 
solutions in $H$ must be regarded as distinct from the others, and in order to single out a unique solution, it must be specified at {em each and every point of the manifold}. 

In particular, no well-posed initial-value and/or boundary-value problem can be posed for such covariant equations. So they would seem not to be of much use, which is why Einstein and Hilbert initially rejected them. \footnote{ See \cite{Stachel1989} pp. 91-101 or \cite{Stachel2002}, pp. 353-364}
But we know from the example of general relativity that they are. Therefore, there must be a way to evade the hole argument, as Einstein finally realized 
\footnote{For a historical review, see \cite{Stachel1989}. }
One must assume that, at least inside the hole, the points of the manifold 
are {\em not} individuated independently of the fields. 
In the gravitational case, this means that the space-time points are not individuated 
independently of the corresponding physical fields , 
i.e., they have no inherent chrono-geometrical or inertio-gravitational properties or 
relations that do not depend on the presence of the metric tensor field. 
In that case, when we carry - or drag-along the solution, we carry- or drag-along the physically individuating properties and relations of the points. Thus, the carried- or dragged-along solution does not differ {\em physically} from the original 
one. \footnote{The concept of {\em physically the same} is given a precise meaning in ({\em Section 3}}

Put in other words, while the points of the manifold have an inherent {\em quiddity} as elements of 
space-time, they lack {\em haecceity} \footnote{For a discussion of {\em quiddity} and {\em haecceity}, see
\cite{Stachel2004}, pp. 204 and \cite{Stachel2005}}as individualized points of that space-time (``events``) 
unless and until a particular metric field is specified.
\footnote{In the generic case (i.e. no symmetries present), 
the 4-non-vanishing invariants of the Riemann tensor in empty space-times can be 
used to individuate the points of space-time. These are the so-called Kretschmann-Komar 
coordinates (see \cite{Stachel1993}, pp. 155-156)}.  It follows 
that the entire class of diffeomorphically-related solutions to the homogeneous ('empty space') 
field equations corresponds to just one inertio-gravitational field.\footnote{Those not familiar
with the details of the hole argument will find a generalized version, applicable 
to any geometric object field, explained in the {\em Appendix}.}

It is clear that a generalized version of the hole argument can also be applied to regions $H$ of a 
4-manifold $M$, in which the inhomogeneous Einstein equations hold, together with the set of 
dynamical equations obeyed by the non-gravitational matter and fields acting as sources of 
the metric field, provided this set of coupled gravitational and non-gravitational field 
equations has the covariance property (we leave the details to the reader).
\footnote{Those not familiar
with the details of the hole argument will find a generalized version, applicable 
to any geometric object field, discussed in detail in the {\em Appendix}.} In order to 
avoid this version of the hole argument, it must be assumed that the points of this region are 
not individuated unless and until both the gravitational and non-gravitational  in the region 
are specified.\footnote{This does not imply that {\em all} of these fields are 
necessary for such individuation. The values of four independent invariants of the fields will 
suffice to individuate the points of the space-time in the generic case 
(i.e., when no symmetries are present).}  
Again, an {\em entire class} of diffeomorphically related solutions to the coupled equations 
will correspond to one physical gravitational and non-gravitational fields.      

Put into the language of fibered manifolds (discussed in {\em Section 2}), the root of the hole argument lies in the possibility of {\em independent} base-space 
diffeomorphisms and total-space automorphisms. So, a {\em necessary} condition for 
being able to pose the hole argument is the possibility of a base space defined independently of the total space. But this condition is not {\em sufficient}. 

A cross-section of a fiber manifold defines a field of quantities on the base manifold. 
It may define a physical field, by choosing appropriately the fiber manifold. \cite{Crampin}.
The hole argument fails if the points of the base space are not individuated independently 
of a {\em cross-section}.

In {\em Section 3}, we shall generalize the hole argument in two ways:
\footnote{See \cite {MacLane}, pp. 434 : `` Generalization from cases refers to the way in which several specific prior results may be subsumed under a single more general theorem.``} 

1) the dimension of the differentiable base manifold is arbitrary but finite, and

2) the metric together with any other tensor fields will be replaced by an arbitrary 
geometrical object field.\footnote{The geometrical object field will not be assumed irreducible, 
in order that one can combine several irreducible fields into one such field; 
e.g., in the case of Einstein-Maxwell equations, the metric $(g_{\mu \nu}$ and the 4-potential  $A_{\mu \nu})$ taken together do not form an irreducible geometric object.} 

The generalization is to the category of fibered manifolds, 
which includes such important special cases as fibre bundles, their jet prolongations, 
and connections on bundles \cite{Saunders}, \cite{Leon}. 
This permits the treatment of the physically important case 
of gauge field theories\cite{Darling}, as well as first order, Palatini-type formulations of general 
relativity based on an independent affine connection as well as a metric field.
\footnote{ See for example \cite{Wald}, {\em Appendix E}, pp. 454-455.}  
In all such cases, the underlying manifold $M$ is taken as the base space of the appropriate 
fibered manifold. Stachel in \cite {Stachel1986} gives a similar generalization of the hole 
argument, based on the concept of fibered manifolds,  
with the primary aim of making it clear that the argument is 
coordinate-independent.\footnote{In \cite{ Stachel1986} the treatment is based on fibered spaces
and their automorphisms as in \cite{Hermann}. The paper is written in the language of fiber bundles, 
but it actually uses only the properties of a fibered manifold, and defines a geometric object 
as a cross section of such a manifold. (To facilitate comparison, we use the same notation here
 as in that paper.) Earman \cite{Earman1989}, pp. 158-159, follows \cite{Schouten}, pp. 67-68, in giving a coordinate-dependent 
definition of geometric object fields, and implies that a coordinate-independent definition would be very difficult. 
Earman cites Stachel \cite{Stachel1986} in another context, without noting that it gives such definition. }\label{Note16} 

However, it does not take what seems the obvious next step elimination of an independent base 
space, which is outlined in the remainder of the {\em Introduction} and discussed in detail in 
{\em Section 3}.

A fibered manifold consists of the triple: total space, base space, 
and a projection from total space into base space. The inverse of the projection operation 
defines the fibers of the total space. If all the fibers are isomorphic, one has a fiber bundle. 
A natural fiber bundle represents some type of geometric object field, and a 
cross-section of it represents a particular field of the given type. 
({\em Section 2.1})

Both the total space and the base space of a fiber manifold are subject to diffeomorphisms, with the proviso that the total space diffeomorphisms are required to be fiber-preserving. 

A theory is a mathematical choice of fibered manifolds. A type of geometric object is the most general type of fibered manifold that preserve the unique representation of the base diffeomorphisms. The representation group is the group of fiber-preserving automorphisms.

For theories for which the hole argument is not valid, we can  make a move to block its formulation. 
Our proposal is to avoid the hole argument, by  re-formulating the relation between the total and base space in such a way that we cannot even formulate the hole argument.
The idea is to start with a fibered total space, and
{\em define} the base space as 
the quotient space of the total space by the fibration ({\em Section 3.2}). 
Now the base space is just derived ( quotient space) and not a primary constituent of the fiber manifold from the total space by the fibration (relation).
There is no {\em independent} base space (no fibered total space, no base space); and it 
is {\em impossible} to carry out a fiber-preserving diffeomorphism without inducing the corresponding 
diffeomorphism of the base space. So, one cannot drag along a cross-section without dragging 
along the points of the base space. The hole argument cannot even be formulated! Therefore this 
seems the ideal way to treat situations, such as general relativity or any generally-covariant 
set of field equations, in which we do not want independently individuated points of the 
base space.

{\em Section 4} sets out the essence of the hole argument for sets. The category of sets is reached 
through a process of abstraction by deletion from the category of differentiable manifolds.
\footnote{See \cite{MacLane}, p. 436: ``Abstraction by deletion is a straightforward process: One carefully omits parts of the data describing the
 mathematical concept in question to obtain the more {\em abstract} concept.``}  
If we abstract from the differentiability, but keep continuity, we reach the category of 
topological manifolds; and if we abstract from continuity, we are left with bare sets. 
The same process of abstraction enlarges the automorphism group of $M$ from the diffeomorphism 
group to the homeomorphism group of the resulting topological manifold; and then to $Sym M$, the symmetric group of the resulting set $M$, also called the group of all permutations (bijections maps)
of $M$. 
\footnote{When a set contains an infinite number of elements, the term "permutation" 
is sometimes limited to the case when only a finite number of elements are permuted, 
which would correspond to a similar abstraction from the concept of local diffeomorphism, 
discussed below.} Invariance under the group of diffeomorphisms 
becomes invariance under the homeomorphism group, and then invariance under the symmetric group.\footnote{ See, for example, \cite{Smith} pp. 78,
or \cite{Neumann}, pp. 9.}
 Ultimately, it may prove important for 
physics to consider the intermediate abstraction to homeomorphisms of topological manifolds; 
but in this paper we shall consider only the direct abstraction from diffeomorphisms of 
differentiable manifolds to permutations of sets.
\footnote{In the language of category theory (see e.g., \cite{Adamek}, pp. 22), 
one applies forgetful functor from the category of differentiable manifolds 
with diffeomorphisms to the category of sets with permutations.} In this process of abstraction, 
a fibered manifold representing some type of geometrical object under the diffeomorphism group 
becomes a $G$-set\footnote{Some books also use the name $G$-space for a $G$-set, e.g. \cite{Neumann} pp 32.} with a $G$-invariant equivalence relation, 
where $G$ is the transformation group of the set. \cite{Neumann} 
One can formulate a version of the hole argument for such a $G$-sets that is not merely analogous 
in a general sense to the geometrical-object version, but that can be derived from the latter 
by the application of a forgetful functor.

{\em Section 6} is concerned with the importance of special case of {\em relations}, and shows that the set of all relations between the elements of a set is an example of a $G$-set with a $G$-invariant equivalence relation.

In {\em Appendix} it is described in detail how a fiber automorphism of a tensor bundle is constructed by lifting a local diffeomorphism of $M$ with the help of a linear frame field in $M$ and thereby connect up with the usual definition of a tensor.


\section{Geometrical Object Fields}

We will start by describing the basic structures upon which our study of the hole argument for geometric objects is based.

\subsection{Fibered Manifolds and Geometric Objects}

A {\em (global) diffeomorphism} between two manifolds $M$ and $N$ of the same dimension $n$ 
is a differentiable mapping $f :M\to N$, such that
$f^{-1}$ exists and is differentiable (of the same order as $f$).
If there exists such of diffeomorphism, then $M$ and $N$ are said to be {\em diffeomorphic}.
We shall be concerned primarily with automorphisms of $M$, i.e., diffeomorphisms 
$f :M\to M$ of $M$ upon itself.  \cite{Crampin}
 
A mapping $\phi :M\to N$ between two manifolds $M$ and $N$ of the same dimension is 
a {\em local diffeomorphism} if each point of $M$ has an open neighborhood $U\subseteq  M$ such that 
the restricted mapping  $\phi_{U} :U\to \phi(U)$ is a diffeomorphism.  

Local diffeomorphism\footnote{If the global topology is not fixed, then we want to use local diffeomorphisms instead of global diffeomorphisms} is a weaker concept than diffeomorphism. \footnote{Local diffeomorphisms are not necessarily even surjective or injective mappings.}

A mapping $\pi:E\to M$  is a {\em submersion at a point} $x$ in $E$, 
if the rank of the linear mapping  $d_{x}\pi:T_{x}E\to T_{\pi(x)}M$ is equal to the dimension of the base manifold $M$.
It is a {\em submersion},  if it is a submersion at each point $x$ of $E$.

A {\em fibered manifold} is a triple $(E, \pi, M)$ , where $E$ and $M$ are manifolds and $\pi$  
is a  surjective submersion. $E$ is called the {\em total space},  $M$  is the {\em base space}
and $\pi$ the {\em projection}.

Every fibered manifold admits {\em local cross-sections} i.e., smooth maps 
 $U\stackrel{\sigma }{\longrightarrow }E$ 
defined on {\em open sets of the base space} $U\subseteq M$  such that $\pi \circ \sigma =id_{M}$ .  
This means that, for every point $x$ of $E$, there exists an open neighborhood $U$ of $\pi= \pi(x)$ in $M$ and local cross-section $U$, $\sigma_{x}:U\rightarrow E$.

The weaker concept of local diffeomorphism is suited for use with local sections. Not 
all  fibered  manifolds admit {\em global sections}, e.g., a principal bundle admits a global 
section if and only if it is trivial, while a vector bundle admits an infinite number of 
global sections. \cite{KMS1993}

For each $p\in M$, the set $\pi^{-1}(p)=E_{p}$ denotes the {\em fiber over} $p$. 
The fibers are themselves differentiable manifolds (actually submanifolds of the total space $E$), 
but they are not necessarily isomorphic. If all the fibers are isomorphic to some manifold $F$ 
({\em typical or standard fiber}), then we  have a {\em fibre bundle} \footnote{We shall carry out all the proofs 
for the general case of fibered manifolds. However, from the definition of a type of geometric object 
(natural bundle), it can be shown (see \cite{KMS1993}) that for each $n$-dim manifold $M$, the functor $F$ actually 
induces a fibre bundle structure on the value $FM$}\cite{Husemoller}.

A fibered (manifold) morphism  $\phi$ (over $\bar{\phi}$) between two fibered manifolds $(E\stackrel{\pi_{M}}{\longrightarrow }M )$ and
$(E'\stackrel{\pi_{M'}}{\longrightarrow }M' )$
is a differentiable map 
$\phi:E \rightarrow E'$ which map each fiber of $\pi$ into a fiber of $\pi '$. 
The map $\phi$  between total spaces $E$ and $E'$ determines uniquely the map $\bar{\phi}$ between the base spaces \cite{KMS1993}.
The map $\bar{\phi}$  is the {\em projection} of $\phi$. If the two fibered manifolds are 
identical, i.e.,$(E\stackrel{\pi_{M}}{\longrightarrow }M )= (E'\stackrel{\pi'_{M'}}{\longrightarrow }M' )$ then $\phi : E\to E$ an {\em automorphism} of $E$, that defines a local diffeomorphism 
$\bar{\phi}_M$ of $M$, which we shall refer to as the {\em projection diffeomorphism} of $\phi$.

A type of geometric object is the general class to which a particular geometric object belongs. Tensors are a proper subclass of all geometric objects and so, for example, metric tensors are a type of geometric object: symmetric tensors of rank $(0,2)$. The Schwarzschild metric is an example of a particular geometric object within this type.

In modern differential geometry, a {\em type of geometric object}
is defined as a {\em natural bundle} over the category of smooth manifolds. 
We will follow here the coordinate-independent  definition given in \cite{KMS1993}.
 	
Let ${\mathcal M f_n}$ be the category of smooth $n$-dimensional manifolds, i.e. 
the category whose objects are $n$-dimensional manifolds and whose morphisms are local 
diffeomorphisms and let ${\mathcal FM}$ be the category whose objects are fibred manifolds and 
whose morphisms are fibre-preserving morphisms.

A {\em type of geometric objects} or {\em natural bundle} is defined as a 
covariant functor $\mathtt{F}$  of ${\mathcal M f_n}$ into ${\mathcal FM}$, i.e.,
$\mathtt{F}$ is a rule such that:

\begin{enumerate}
\item Each $n$-dimensional manifold $M$(object in in the category ${\mathcal M f_n}$) is transformed into a fibered manifold 
$FM \stackrel{\pi_M}{\longrightarrow }M$ (object in in the category ${\mathcal FM}$)

\item Each local diffeomorphism  $\phi :M\to N$ between two n-dim manifolds $M$ and $N$
(morphism between two objects in the category ${\mathcal M f_n}$) is transformed into 
a fibered preserving morphism  $F\phi$, $\pi_{N}\circ F\varphi=\varphi \circ \pi_{M}$(
morphism in the category ${\mathcal FM}$)

\item If $\phi :M\to N$ and $\psi  :N\to P$ are two local diffeomorphisms, then 
$F(\psi\circ \phi)=F\psi\circ F\phi$ is a fiber-preserving morphism over $\psi\circ \phi:M\rightarrow P$

\item  For every open set $U\subseteq M$, the inclusion map $i:U\rightarrow M$ is transformed 
into the inclusion map $Fi:FU=\pi^{-1}_{M}(U)\rightarrow FM$

\end{enumerate}

A {\em geometric object} is a defined as a(local) cross-section of some type of geometric object $F$.  
\footnote{In what follows we shall not distinguish notationally between a natural bundle 
$F$, the fibered manifold $FM\to M$ or the total space $FM$.}

An example of natural bundle is the tangent bundle $TM$. \footnote{In the case of a tangent 
bundle $TM\stackrel{\tau_M}{\longrightarrow }M$,  $(x^{a}, y^{i})$ local fiber coordinates.
Then, to any base coordinate change $x^{a}\longmapsto \bar{x}^{b}=\phi^{b}(x^{a})$ it is uniquely 
determined fibered coordinate change $y^{i}\longmapsto \bar{y}^{j}=\Phi^{j}(x^{a}, y^{i})$.} 
Also, the cotangent bundle $(T^{*}M)$ and any tensor bundle, e.g., the bundle of all 
pseudo-symmetric $(0,2)$ forms over a smooth manifold, is a natural vector bundle. 
The linear frame bundle $(LM\rightarrow M)$ 
, which is the principal bundle associated to the tangent bundle, is a natural bundle. 
\footnote{See for further discussions of frame bundles at end of {\em Section 2} and the {\em Appendix}} 

But in general, principal bundles are not natural bundles. There is a generalization of the notion of a natural bundle to the case of principal bundles, called " gauge-natural bundle " ( e.g \cite{Matteucii} and \cite{KMS1993}, {\em chap. XII})

\subsection{Internal Automorphisms and Gauge Transformations}

Many physical theories are gauge invariant.  In such theories there will often be a 
class of cross-sections of a suitable fibered manifold, each member of which represents the {\em same} physical model\footnote{See {\em Section 3.1} for a formal discusion of physical models} of the theory. 
One member of the class is related to another one by a transformation belonging to $Gau(E)$, the 
{\em gauge group}  . 

A {\em gauge transformation} $\phi \in Gau(E)$ is a fiber( manifold) automorphism
$\phi : E\to E$ such that the projection diffeomorphism is the identity base diffeomorphism
$id_{M}:M\to M$. In other words, $\phi$ leaves the points of the base space $M$ unaltered and 
takes points of the same fiber into each other, i.e., $\phi(E_{x})=E_{x}$ for all $x\in E$.
Such an automorphism is also called an {\em internal automorphism} and hence is denoted $\phi_{int}$.

Important example of fibered manifolds with internal automorphisms 
are $G$-bundles\footnote{See \cite{KMS1993}, pp. 86-87} with
a left-action of a Lie group $G$ on each fiber; principle fiber bundles, that are 
$G$-bundles, in which each fiber is itself diffeomorphic to $G$ (as a manifold, but the fibers fail to 
be groups).
\footnote{In this case, there are two main types of gauge transformation, 
which locally are equivalent. These are {\em atlas transformations} and {\em associated principal 
morphisms}. See \cite{Sardanashvily}, pp 38-45 }

\subsection{External Automorphisms}

An {\em external automorphism} of a fibered manifold $(E,\pi, M)$ is a fiber 
automorphism, i.e., $\phi : E\to E$ that  projects over an arbitrary (local)
diffeomorphism $\phi_M$ of $M$ and such that $\pi\circ\phi=\phi_{M}\circ{\pi}$ .

The local diffeomorphism $\phi_M$ of the base manifold is uniquely associated with 
every fibered manifold automorphism $\phi$. \cite{KMS1993} 

Now, let us look at the inverse problem: suppose we start  with a local diffeomorphism 
$\phi_M$  of $M$ that carries a point $x\in M $ into point  $y=\phi_{M}(x)$ of $M$.
What are the fiber automorphisms associated with it?
First, the identity diffeomorphism $id_M$ may have the entire class of internal 
automorphisms $\{\phi_{int}\}$ associated with it. 
For other diffeomorphisms $\phi_{M} \neq id_{M}$, the only natural restriction on a fibered  manifold automorphism $\phi:E\to E$ is that it should {\em preserve the relation between  points and the fibers over them}:  
if  $p$ and $q$ belong to some small open set $U\subseteq M$ and $\phi_{M}(p)=q$ the fiber automorphism 
$\phi$ should carry the fiber $\pi^{-1}(p)$  into the fiber $\pi^{-1}(q)$, 
both in the open subset $\pi^{-1}(U)\subseteq E$.\footnote{Since the automorphism is local, 
the two fibers cannot fail to be isomorphic}
Generally, for any $\phi_M$ automorphism of a fiber manifold,  there will be many external 
automorphisms $\phi$ that meet this requirement. Indeed, given one such $\phi$, 
then $\phi_{int}\circ\phi$  will also meet the requirement for all $\phi_{int}$ . 

We may further restrict the external automorphism by demanding that they form a 
{\em representation} 
of the local diffeomorphism group. That is: 

a) If $\phi$ is a fibered manifold automorphism corresponding to the local diffeomorphism $\phi_M$ , then $\phi^{-1}$ corresponds to $\phi_{M}^{-1}$ .

b) If two local diffeomorphism $\phi_{M'}$ and $\phi_{M''}$ 
are such that $\phi_{M}=\phi_{M'}\circ \phi_{M''}$, and if $\phi '$ corresponds to $\phi_{M'}$ and $\phi ''$ corresponds to $\phi_{M''}$, then $\phi=\phi '\circ \phi ''$ corresponds to $\phi_{M}$. 

Because of possibility of internal automorphisms, this condition is still not enough to always associate a unique fibered manifold automorphism with each local diffeomorphism. But 
we can associate the unique equivalence class $\{\phi\}$ of all automorphisms $\phi$ 
that correspond to the same $\phi_M$. 
There is an equivalence relation (i.e. reflexive, symmetric and transitive)of fiber 
automorphisms, such that two fibre automorphisms $\phi$ and $\phi'$ are {\em equivalent}, if there exists an internal automorphism $\phi_{int}$  such that $\phi'= \phi\circ\phi_{int}$. 
This equivalence relation therefore partition the automorphisms into equivalence classes. 

These equivalence classes provide a representation of the local diffeomorphism group.
\footnote{Following \cite{Hermann}, Stachel in \cite{Stachel1986} calls 
the pair consisting of $\phi$ and $\phi_M$  a fiber space automorphism.}  
The importance of this requirement is that, at the abstract level of the category of fibered 
manifolds, it represents the translation of the requirement 
that a geometric object field transform uniquely under a diffeomorphism. A type of geometric object can therefore be defined as a fibered manifold  $(E\stackrel{\pi_{M}}{\longrightarrow }M )$
with a unique equivalence class $\{\phi\}$ associated with every $\phi_M$.

Physical models of a theory are identified with classes  $\{\sigma[p] \}$ of local sections under the internal automorphisms.  
Under an external fibered (manifold) automorphism $\phi:E\to E$, a local section equivalence class 
$\{\sigma[p] \}$ is carried into another (unique) local section equivalence class, 
symbolized by $\{\phi^{*}\sigma[p]\}$. \footnote{We will give a formal definition of  $\phi^{*}\sigma[p]$ in {\em Sec} 3.1}
That is, $\{\phi^{*}\sigma[p]\}$  is the local section equivalence class that 
results from $\{\sigma[p]\}$  if we carry out the fibered manifold automorphism equivalence 
class $\phi$ without any automorphism of the base space, i.e., 
while leaving the points of the base manifold unchanged. To simplify the language and notation, from now on all reference to equivalence classes  of automorphisms and 
cross-sections will be omitted, being understood that they should  be inserted 
whenever the theory in question has an internal gauge group.

\subsection{Yang-Mills vs General Relativity}

The definition of the group of external automorphisms leaves open the question of its 
relation to the group of internal automorphisms. 
The two may be entirely independent of each other, as in the case of Yang-Mills theories (including Maxwell's theory), in which the space-time diffeomorphisms are independent of the internal gauge transformations. 

In Yang –Mills theory, the fibered manifold $(E\stackrel{\pi_{M}}{\longrightarrow }M )$ is a principal bundle with a Lie 
group structure $G$ and the gauge transformations are principal automorphisms over $id_M$. 
An external (general) principal automorphism $\phi$ is a $G$-equivariant diffeomorphism  of $E$ on itself, i.e. $\phi:E\to E$ is a diffeomorphism such that $\phi(pg) =\phi(pg)$, for all $g\in G$, and $r_{g}(p)\equiv  pg$ denotes the canonical {\em right action} of $G$ on the total space $E$. 
We denote with $Aut (E)$ the group of all fiber automorphisms of $E$.  
Since $\phi:E\to E$ is fiber preserving, there is a group homomorphism $\mathbf{pr}$ from  
$Aut(E)$ to the group $ Diff M$ of all diffeomorphisms of $M$ such that 
$Ker \mathbf{pr} = Gau(E)$, where $Ker \mathbf{pr}$ is the kernel of the projection. An external principal automorphism $\phi$  of $E$ is associated with a tangent bundle automorphism $(\phi_{M})^{*}$ of $TM$ that project on $\phi_M$.

It follows that the following short sequence of group homomorphisms:

$$\{e\}\to Gau(E) \stackrel{i}{\longrightarrow } Aut(E) \stackrel{pr}{\longrightarrow } Diff M\to \{e\}$$
is exact.\footnote{A short sequence of group homomorphisms  
$\{e\}\to G_{1} \stackrel{f_{1}}{\longrightarrow } G_{2} \stackrel{f_{2}}{\longrightarrow } G_{3}\to \{e\}$ is called exact, if  $f_{1}$ is injective , 
$f_{2}$ surjective and $Imf_{1}=Kerf_{2}$  (see \cite{Lang}).}
The sequence does not split, since
we cannot define a homomorphism $f:Diff M \to Aut(E)$ such that 
${pr}\circ f=id_{Diff M}$. This is because, as shown above, to each
$\phi_M\in Diff M$ there may be many automorphisms $\phi \in Aut(E)$ that project over $\phi_M$.

Now the case when the fibered manifold $(E\stackrel{\pi_{M}}{\longrightarrow }M )$ belongs to the category of natural bundles (i.e. a type of geometric objects),
this means that any local diffeomorphism $\phi_M$ of $M$ admits a canonical lift 
(to a fiber automorphism $\phi$ of $E$) , treated as a external transformation. 
The structure group is $GL(n, \mathbf{R})$. General Relativity is formulated on such a fiber manifold 
for which  $\mathrm{dim} M = 4$  and $M$ satisfies topological conditions such that $GL(4, \mathbf{R})$ 
is reducible to the Lorentz group $SO(1,3)$.\footnote{The corresponding Higgs filed is a pseudo-riemannian metric. 
A pseudo-riemannian metric and a connection on $LM$ will define a metric-affine gravitational theory \cite{Sardanashvily}}
The associated principal bundle is the bundle  of all linear frames\footnote{see {\em Appendix}}  
in the tangent bundle $TM$ of $M$. The projection map $p$ is the mapping that 
associates to each frame the point in which the tangent space lies and its standard fibre is 
the linear group $GL(n, \mathbf{R})$ . 
In this case, the exact sequence of group homomorphism 
$$\{e\}\to Gau(E) \stackrel{i}{\longrightarrow } Aut(LM) \stackrel{pr}{\longrightarrow } Diff M\to \{e\}$$
splits. This means that the group $Aut(LM)$  can be written as the direct sum of the 
groups $Gau(LM)$ and $Diff M$.

To prove this we notice that every local diffeomorphism $\phi_M$ of $M$ {\em has lift to a 
diffeomorphism} $\phi_{LM}$ of $LM$ , i.e.,  $\phi_{LM}=f(\phi_{M})$. 
But this condition is equivalent with $f$ is a diffeomorphism 
of $LM$ which preserves the {\em canonical soldering form }  
$\theta =\partial _{i}\otimes dx^{i}$ on $M$, i.e., $f^{*}\theta =\theta $, 
which is true. \cite{Trautman}

On the other hand every principal automorphism  $\phi_{LM}$  will induce an associated bundle automorphisms $\phi$ of $E$
\cite{Sardanashvily}. As in the case of general relativity, the internal and external 
automorphisms are closely related, a space-time diffeomorphism induces an affine transformation of 
the basis vectors at each point of the frame bundle (see the {\em Appendix}). Generalizing the 
terminology often applied to this case, one may say that in such cases the internal automorphisms 
are {\em soldered} to the external automorphisms.

\section{The Hole Argument for Geometric Objects}

\subsection{Formulating the Hole Argument}

We generalize the hole argument for geometric objects as follows. 
A theory is a mathematical choice of fibered manifolds. 
Let $M$ be a finite dimensional manifold and ($E\stackrel{\pi}\to M$) a fiber manifold over $M$. 
A cross-section $\sigma $ of the fibered manifold $E$ defines a field of physical quantities on $M$.
\footnote{In general relativity case, the base space $M$ is a 4-manifold and the total space $E$ is the tensor space of $(2,0)$ pseudo-symmetric tensors and the projection $\pi$ is the map that associate to each $(2,0)$ pseudo-symmetric tensor the base point in $M$.}
A {\em model} consists of a fibered manifold $E$ and a`` global " cross-section 
\footnote{Not all fibered manifolds admit global cross-sections, but every fibered manifold has local cross-sections. If it admits a global cross-section( defined on the whole manifold $M$) then it admits a global cross-section that is the extension of any given local cross-section. If the fibered manifold does not admit a global cross-section, then we can still pick up an {\em atlas} of local cross-sections defined on open manifolds such that they form an open cover of the whole base manifold $M$.} of some type of geometric object $\sigma[p]$.

Let $\phi_M: M\to M$ be a local diffeomorphism on $M$. Then from the definition of a natural bundle, there is a uniquely defined fibered automorphism $\phi: E\to E$ that projects over $\phi_M$.

A fibered automorphism  $\phi$ can be used to transport local cross-sections of $\pi$ to local cross-sections of $\pi$:  if $\sigma[p]$ is a local cross-section of $\pi$ defined on a small open set $U_p$ around $p$, then $\phi^{*}\sigma[\phi_{M}(p)]$ defined by
$\phi^{*}\sigma =\phi\circ\sigma\circ\phi^{-1}_{M}$ is a new local cross-section of $\pi$
defined on the open set $U_{q}=\phi_{M}(U_{p})$ around $q=\phi_{M}(p)$.
It is called the {\em carried-along cross-section} of $\sigma$ by the fibered automorphism $(\phi, \phi_{M})$.

We have two mathematically distinct local cross-sections  $\sigma[p]$ and $\phi^{*}\sigma[q]$, where
$p, q \in M$ such that $q=\phi_{M}(p)$.
Now from the definition of a model it follows that $(E,\sigma[p])$ and 
$(E,\phi^{*}\sigma[\phi_{M}(p)])$ are two {\em elementarily equivalent} models i.e., they share the same model-theoretic properties \footnote{See e.g. \cite{Hodges}, pp 43}. 
The truth values or the probability of the corresponding assertions in each model will always be the same. That is for every assertion about the model $(E, \sigma[p])$, there is a 1:1 corresponding assertion about the model $(E,\phi^{*}\sigma[\phi_{M}(p)])$.
Since it follows from the definition of $\phi^{*}\sigma$  that this 
semantic identity cannot fail to hold \footnote{For general covariant theories, this follows automatically. For theories that are not general covariant, the identity still holds: when you move everything, you move nothing !}
,  we shall refer to it as the {\em trivial identity}.
\footnote{There is a more detailed discussion of the trivial identity in the {\em Appendix}.}

We define a theory $\mathcal T$ based on some geometric object, as a rule for selecting a class of models \footnote{In other words, a theory consists of a  rule for selecting a class of cross-sections of a fibered manifold defining the type of geometric object that the theory deals with, while a model is a choice of a particular cross-section. A model of a theory is a choice of a cross-section obeying a rule.} of that type of geometric object.  \footnote{Here we follow the traditional account, which keeps the base manifold fixed. 
Later, we shall modify this account to allow different local cross sections to correspond to different base manifolds. In most current physical theories, the selection rule is usually based on the solutions to some set 
of differential equations for the type of geometric object; the formulation of the rule therefore 
involves jet prolongations of the fibered manifold 
(see \cite{Hermann}, Chapter I, pp. 1-15 and  \cite{KMS1993}, Chapter IV,  pp. 124-125); 
but here we eschew such, otherwise important, complications in order the highlight the main point.}

Now under an (external) fibered automorphism $\phi:E\to E$, a local cross-section 
$\sigma[p]$ can be carried into another (unique) local cross-section, symbolized by $\phi^{*}\sigma[p]$. 
Intuitively, as shown at the end of {\em Section 2.3}, 
$\phi^{*}\sigma[p]$  is the local cross-section that 
results from $\sigma[p]$  if we carry out the fibered manifold automorphism $\phi$, 
while leaving the points of the base manifold unchanged. 
To define it more precisely, we take the carried-along cross-section 
$\phi^{*}\sigma[\phi_{M}(p)]$ defined as above on $U_q$, then $\phi^{*}\sigma[p]$ is the {\em pull-back} cross-section by the map $\phi_M$ defined by 
$\phi^{*}\sigma[p]:=(\phi^{*}\sigma[\phi_{M}(p)],p)$ on the open set $U_p$. \footnote{Actually $\phi^{*}\sigma[p]$ is a local cross-section of the pull-back fibered manifold 
$\phi^{*}_{M}E\cong E$ and can be seen as the union of all fibers of $E$ transplanted from $M$ to $M$
using the local diffeomorphism $\phi_M$}

Now, the original hole argument generalized to the case of geometric objects translates in the possibility that under a fiber diffeomorphism $(\phi, \phi_{M})$ the two models 
$(E,\sigma[p])$ and $(E,\phi^{*}(\sigma)[p])$ do not represent the same possibility. 

If  we impose the following {\em covariance} requirement\footnote{Our approach may be contrasted with that expoused in \cite{Earman}, which calls {\em generally covariant} theories that we call {\em covariant}. Our definition of {\em generally covariant} is given below.}
on $\mathcal T$: if $(E,\sigma [p])$ is a model of $\mathcal T$, then so is $(E,\phi^{*}\sigma[p])$, for all local diffeomorphisms $\phi_M$, but as we explained above, they are {\em not} trivially identical. It matters - or rather it may matter - if we permute the fibers {\em without} permuting the base points or, equivalently, permute the base points without permuting the fibers. 

In particular, this will be the case if the models are defined as solutions to a set of covariant 
field equations for the $\sigma [p]$ fields.\footnote{Indeed, we may take this as the definition of 
a set of covariant differential equations for the geometric object field. 
The jet prolongation of the fibered manifold will then have to be used to 
formulate these equations. In view of the widespread confusion about 
the meaning of covariance, it is worth emphasizing that this definition is {\em coordinate-free}.}  
Two models related by such a local diffeomorphism are called 
{\em diffeomorphism-equivalent}. This relation is clearly an equivalence relation, so it divides all models of $\mathcal T$ into diffeomorphically-equivalent classes.

The hole argument hinges on the answer to the following question. 
Is it possible to pick out a {\em unique} model within an equivalence class by specifying $\sigma [p]$
everywhere on $M$ except on some open  submanifold $H\subseteq M$ (the {\em hole}),  i.e., 
on $M -H$? 

If $(E,\sigma [p])$ and $(E,\phi^{*}\sigma[p])$ are not elementarily equivalent 
models for all $\phi_M$ (except the identity diffeomorphism $id_M$, of course), then the answer is "no". 
\footnote{Remember that, by definition of a covariant theory, all of the $\phi^{*}\sigma$
 are models of $\mathcal T$ if one is.}
For we can then pick any local diffeomorphism 
$\phi_{M-H}$ that is equal to the identity diffeomorphism on $M -H$, but differs from the identity 
diffeomorphism on $H$. Then $(E,\sigma [p])$ and $(E,\phi^{*}\sigma[p])$
will be two different models that agree on $M - H$ but differ on $H$, 
and are therefore non-identical.\footnote{Of course, they belong to the same equivalence class, 
but the point is that this does not {\em automatically} make them elementarily equivalent}  
In this case, no conditions imposed on $\sigma [p]$ on $M -H$ can serve 
to fix a unique model on $H$, no matter how small the hole is; the only way to specify such a model 
uniquely, is to specify $\sigma [p]$ everywhere on $M$. 

This is the hole argument for covariant theories - or rather, against them if we require 
a theory that specifies a unique model under the appropriate assumptions.\footnote{We shall not here go 
into the well-known reasons why the uniqueness requirement 
seems a reasonable one to demand of a physical theory.}
Why does it work? Its validity  depends crucially on the assumption 
that the distinction between the points of the manifold, which we call their haecceity,
\cite{Stachel2005} is {\em independent} of the specification of a particular model of the theory $\mathcal T$, that is 
(assuming $M$ given) is {\em independent} of the specification of $\sigma [p]$ .

If the individuation of the points of $M$  {\em does} depend entirely on the model, then
we have no grounds for asserting that $(E,\sigma [p])$ and $(E,\phi^{*}\sigma[p])$
 represent different models.
 For using the trivial identity with $\phi_M$ replaced by $(\phi_M)^{-1}$, we see that 
$(E,\phi^{*}\sigma[p])$ and $(E,\sigma[\phi^{-1}_{M}(p)])$
 are identical; and, given the lack of any model-independent distinction between the points of 
the manifold, no distinction can be made between the models $(E,\sigma [p])$ and 
$(E,\sigma[\phi^{-1}_{M}(p)])$, 
so they are semantically the same model.  Conversely, if $(E,\sigma [p])$ and  
$(E,\phi^{*}\sigma [p])$ 
are identical models for all $\phi_M$, then the hole argument clearly fails; 
and it follows that (as far as concerns the covariant 
theory $\mathcal T$ under consideration)\footnote{One can easily think of cases where the distinction can be made on other grounds. For example, in describing houses built out of cards, two houses will be regarded as structurally identical in structure even if built out of cards 
that can be distinguished independently of their position in the card house.}  
the points of $M$ must be entirely unindividuated before a model $\sigma [p]$ is 
introduced. 
All relevant distinctions between these points must be consequences of the choice of
$\sigma [p]$.  We call such a theory {\em generally covariant}.

\subsection{Blocking the Hole Argument}

In {\em Section 2.1} we defined a fibered manifold as a triplet $(E,\pi,M)$, where $\pi$ is a surjective submersion between the two manifolds $E$ and $M$. There is an equivalent relation $\rho$ on $E$: $\rho(a,b)$ iff $a$ and $b$ are in the same fiber.

Another way to define a fiber manifold is to start with a differentiable manifold $E$ and an equivalence  relation $\rho $ on $E$ such that the quotient space $M = E/\rho $ is a differentiable manifold (of same order of differentiability) and the projection 
$\pi: E\to M=E/\rho$ is differentiable map of 
$\mathrm{rank}\pi = \mathrm{dim}E- \mathrm{dim}M$.{\footnote{ see \cite{Lichnerowicz}. Also in 
\cite{Iftime} an equivalence relation on base manifold determines a fibration of the base manifold $M$ by a space-time foliation}

The two definitions are equivalent, but it will be advantageous to use the second one here 
because, as we shall see below, it will be impossible to carry out a 
fiber automorphism of $E$ without carrying out the corresponding diffeomorphism of the quotient space
$E/\rho$, and vice versa.  

Indeed, suppose we start with a fiber manifold given by a triplet $(E,\pi, M)$ for which the cross-sections represent  physical fields (geometric objects) and suppose there are additional structure on the manifolds. For example the Minkowski metric on $M$ and electromagnetic fields on $E$. If we take a cross section that is a solution to Maxwell's equations in Minkowski space. If we carry out a fiber automorphism that does not preserve the Minkowski metric i.e., one whose projection on $M$ is a diffeomorphism that does not belong to the inhomogeneous Lorentz group, then the drag- along of the solution cross section will {\em not} be a solution of Maxwell's equations in Minkowski space-time.
Our proposal is to avoid the hole argument, by re-formulating the relation between 
the total and base space in such a way that we cannot even formulate the hole argument. We will want to define the fiber manifold as a quotient projection by an equivalence relation $\rho$, so that 
$E/\rho$, the base space 
\footnote{This is similar to what is often done in defining a principal fiber bundle (see \cite{Kobayashi} pp. 50}
 defines a fai). The definition of a principle bundle requires not only that the fibers be isomorphic 
to each other (which is all that is required of a $G$-bundle structure), but that they be 
isomorphic to some Lie group $G$. Then "$M$ is defined the quotient space of (the total space) $P$
by the equivalence relation induced by $G$, $M = P/G$ "}
is defined in terms of the total space $E$ 
\footnote{In fiber bundles case, the manifold structure of the total space is determined uniquely by the manifold structure on the base and the typical fiber( see \cite{Saunders})}.
In such cases,  $E/\rho$ and $M$ are diffeomorphic manifolds, but they are not identical, since $M$ posseses additional structures.

To do this, we introduce a partition $E^*$ of the total space $E$ into disjoint subspaces, 
the union of which is $E$.\footnote{Lawvere and Schanuel \cite{Lawvere}, pp. 82 
use the word {\em sorting} in general, for such {\em partitions} if {\em no sort is empty}. }  
We assume further that all subspaces of $E^*$ are isomorphic. 
Now let $\pi: E\to E^*$ be the surjective map that carries a point of $E$ into an element of $E^*$. 
If $E$ is a topological space, this mapping induces a topology on $E^*$, based on the following 
definition of open sets of $E^*$. A subset $U$ of $E^*$ is open if $\pi^{-1}(U)$
is an open set of $E$. 
$E^*$ together with this topology is the quotient space of $E$. 
$E^*$ defines an equivalence relation $\rho$ on $E$: two elements of $E$ are equivalent if and only if they belong to the same element of $E^*$. One then speaks of the quotient space with respect 
to this equivalence relation: $E^{*}= E/\rho$. 
Alternatively we could start with an equivalence relation, and define $E^*$ 
as the set of equivalence classes with respect to this relation: 
$E^{*}= E/\rho$. We can then define the quotient space of $E$ as above.\footnote{A quotient space can be defined for any topological space and an equivalence 
relation on it (see, e.g., \cite{Iyagana} pp. 1282, and \cite{Munkres}, pp. 139). }

Quotient spaces of topological spaces are not always well-behaved. 
For example, $E^*$ may not  be Hausdorff, even if $E$ is. So, if $E$ is a 
differentiable manifold, it is by no means the case that the quotient space
 of $E$ by an arbitrary partition (or corresponding equivalence relation $\rho $) 
will be a differentiable manifold.
\footnote{For a detailed discussion of quotient manifolds see \cite{Bricknell}, pp. 84-108.}
Rather than go into this question here, we shall simply assume that the partition 
(or corresponding equivalence relation) is such that the quotient space is a differentiable 
manifold. In practice, one can just start from a fibered manifold, throw away the base space, 
and define the equivalence classes of the total space as its fibers. 
The quotient space will then always exist and be diffeomorphic to the original base space.
Now the base space is just {\em derived} (quotient) and not a primary constituent of the fibered manifold 
from the total space $E$ by the relation $\rho $.
But now, by the definition of a quotient space, it is impossible to carry out a 
fiber automorphism of $E$ without carrying out the corresponding diffeomorphism of the quotient space
$E^*$, and vice versa. 
The points of $E^*$ are now necessarily individuated 
(to the extent that they are) only by their place in the total space. 
Indeed, we can go further: if there is no total space, there is no quotient space.

This realizes mathematically one of Einstein's intuitions about general relativity. 
He insisted that, without the metric field (here the total space), there should not be
a bare manifold, as in the usual mathematical formulations of general relativity, but - 
literally nothing: 

``On the basis of the general theory of relativity, space as opposed to 
{\em what fills space} ... has no separate existence … 
If we imagine the gravitational field, i.e., the functions $g_{ik}$ 
to be removed, there does not remain a space of the type (of special relativity), 
but absolutely nothing, and also not {\em topological space}. ... 
There is no such thing as an empty space, i.e., a space without field. 
Space-time does not claim existence on its own, but only as a structural quality of the 
field " (Albert Einstein \cite{Stachel1986}, pp. 1860).

We believe that the approach just outlined is the best way to introduce the 
mathematical formalism needed for general relativity, both logically and pedagogically. 
\footnote{Of course suitably simplified for pedagogical purposes}
Careful accounts of the theory traditionally \footnote{see e.g. \cite{Choquet},]\cite{Crampin}, 
\cite{Dubrovin}} start with a 
bare manifold $M$, and then introduce various fields on it. But in some cases, this leads to the 
erroneous impression that, having introduced $M$,  we are already  dealing with a space-time. 
If we start with a fibered manifold representing the metric and 
perhaps other physical fields, and then introduce $M$ as a quotient space, it becomes 
difficult to forget that, in general relativity, space-time is no more than 
{\em ``a structural quality of the field."}

\subsection{Lie Groups versus Diffeomorphisms}

At this point, it is instructive to contrast
the situation in general relativity with that 
in pre-general relativistic theories. The move from fibered manifold to quotient 
space could also be made in the case of a finite-parameter Lie group, such as the 
inhomogeneous Galileo or inhomogeneous Lorentz(Poincar{\'e}) symmetry groups needed in pre-general-relativistic 
Newtonian or special-relativistic space-time theories. However, in these cases there 
is no need to do so in order to block the hole argument: Finite-parameter symmetry 
groups have a sufficiently {\em rigid} structure to block the hole argument even if the base space 
$M$ is defined independently. If an element of the group is fixed at all points of $M-H$ (outside the hole), then it is fixed at all points of $H$ ( inside the hole).

Thus, in pre-general-relativistic space-time theories, one may adopt either of the 
two points of view about the individuation of the points of $M$. Loosely speaking, 
we may say that these theories are compatible with either an absolute or a relational 
approach to space-time. It is only general relativity that forces us to adopt the 
relational point of view if we want to avoid the hole argument.

\subsection{From Local to Global}

The move to the quotient space also enables one to remove another outstanding difficulty in the 
usual mathematical formulations of general relativity. \cite{Stachel1987} 

In solving the field equation of general relativity, one does not start from 
a prescribed global manifold; rather, one solves the equations on some open set 
({\em coordinate patch}), and then looks for the maximal extension of this solution on, 
subject to some criteria for this extension, e.g., null and/or timelike
geodesic completeness \cite{Stachel1987}. The situation is rather analogous to that in 
complex variable theory, in which one can define an analytic function or a solution of the Cauchy-Riemann equations on some open set, 
and then extends this function to the Riemann sheet that represents its maximal analytic extension. This procedure can be 
formulated more rigorously in terms of sheaves of germs of holomorphic functions 
(see \cite{MacLane}, pp. 351-353).    

Rather than starting with a fixed total space $E$, we can start 
with some open manifold $U$, which is ultimately to represent a submanifold 
of a total space yet to be determined; and a relation $\rho(U)$ on it that serves 
to define the quotient space $U^{*}= U/\rho$. To define general covariance and general covariance locally we need only local diffeomorphisms on $U$ and $U^*$. 
After  finding a solution to the field equations on $U$, i.e., stipulating a cross-section of $U$:
 $\sigma(U)$, we look for the maximal extension $E_ {\sigma}^*$ of $U^*$ and
 $\sigma(E_ {\sigma}^*)$ of $\sigma(U)$ subject to the condition 
that the latter be a solution to the field equations on all of $E_ {\sigma}^*$, 
and whatever additional criteria we may impose to define a maximal extension.

In general, the $E_ {\sigma}^*$ for different $\sigma$ will not be diffeomorphic : 
there need be no single base space that fits all solutions. 
The covariance of the field equations will guarantee that all members of the
 diffeomorphically-invariant equivalence class of solutions that contains $\sigma(E_ {\sigma}^*)$, 
will also be cross sections of $E_ {\sigma}^*$. 
Hopefully, this program can also be formulated more rigorously in terms of sheaves of 
germs of the appropriate geometric objects.\footnote{Haag has already done some of the work for 
quantum field theories (see \cite{Haag}, pp. 326-328}

\section{Categories, Fibered Sets, and $G$-Spaces}

As noted in the {\em Introduction}, the essence of the hole argument does not 
depend on the continuity or differentiability properties of the manifold. 
To get to this deeper significance one must abstract from the topological 
and differentiable properties of manifolds, leaving only sets. We shall start by defining 
the basic structures upon which our study of hole argument for 
sets will be based.

As usual one gets a clearer idea of the basic 
structure of the argument by formulating it in the language of categories ( see e.g. \cite{Lawvere},
\cite{MacLane1997}, \cite{Dodson})

A general map $X\stackrel{g}\to S$ between two sets is also called {\em sorting, stacking} or {\em fibering} of $X$ into $S$ {\em fibers} (or {\em stacks}).  
A {\em section} for $g$ is a map $S\stackrel{\sigma}\to X$  such that $g\circ\sigma =id_S$.
A {\em retraction} is a sort of inverse operation to a section . 
A retraction  for $g$ is a map $S\stackrel{r}\to X$ such that $r\circ g =id_X$. 
If $g$ is not surjective, some of the fibers of $g$ are empty.  
If there are empty fibers, then the  map $g$ has no section.  
If $X$ and $S$ are finite sets and all the fibers are non-empty, then 
the map $g$  has a section and it is said that $g$ is a \em {partitioning} of $X$ \em {into} $B$ 
{\em fibers}. In what follows we will consider only the case when $g$ is a surjective map.

If $A$ and $B$ are two sets, a map $A\stackrel{f}\to B$ 
is an {\em isomorphism} if it has an inverse $f^{-1}$ that is both a  section and a  retraction.
An {\em automorphism} (or {\em permutation}) of a set $A$ is a bijection $P:A\to A$.
The set of all the permutations of a set $A$ forms a group, $Perm(A)$, called 
the {\em symmetric} (or {\em permutation}) group on $A$. 
An object in the category of permutations consists of a set $A$ together with a given 
permutation (automorphism) $A\stackrel{\alpha}\to A$. We denote it by $A^{\alpha} $.
A map between two objects $A^{\alpha} $ to $B^{\beta}$ is a map $A\stackrel{f}\to B$ , 
which preserves the given automorphisms $\alpha $ and $\beta $, i.e., 
$f\circ\alpha =\beta \circ f$.
The composition of two maps $f$ and $g$ can be done by composing them as maps of sets that 
preserves the given automorphism, i.e. if $f: A^{\alpha} \to B^{\beta}$ and 
$g: B^{\beta } \to C^{\gamma } $ then $g\circ {f}: A^{\alpha  } \to C^{\gamma } $ such that
(see \cite{Lawvere}, \cite{MacLane1997}):
$$(g\circ{f})\circ{\alpha }= g\circ{(f}\circ{\alpha})=g\circ{(\beta} \circ {f})\\
=(g\circ{\beta)} \circ {f}=\beta \circ{(g\circ {f})}$$

Returning to a fibering $X\stackrel{g}\to S$ , a fiber automorphism $X^{\alpha}$ of the 
set $X$ is a permutation $\alpha $ of  $X$ which preserves the fibers of $g$.
 A fiber automorphism  $X^{\alpha}$ is naturally associated with a base automorphism $S^{\bar{\alpha}}$, where the base projection of $\alpha $ is a permutation $\bar{\alpha }:S\to S$ of base set $S$.

We are interested in the inverse problem: Given a base automorphism $S^{\bar{\alpha}}$,
how can we define a fiber automorphism $X^{\alpha}$,  such that $\bar{\alpha }$ to be the projection of the permutation $\alpha $ ? 

As in the manifold case we restrict the fiber automorphisms: they must form a 
{\em representation} \footnote{The term "representation" of a group is used in a more general sense to mean a homomorphism from the group to the automorphism group of the object. If the object is a vector space then we have a {\em linear representation}. Some books also use for this the word {\em realization} while representation is rezerved for a linear representations.}
of the transformation group of base-space automorphisms.
We have seen above how to compose mappings of $S^{\bar{\alpha}}$ ; a similar composition rule that 
respects the fibering shall apply to the $X^{\alpha}$. In analogy to the geometric object case, we postulate that to each $S^{\bar{\alpha}}$ there  corresponds a unique $X^{\alpha}$. 

We close this section with a discussion on an important class of such {\em fibered sets}
called {\em $G$- sets}.
Let $\pi :X\to S$ be a {\em fibered sets} for which the fibers contain the same number of elements.
Let $G$ be a finite group. 

An {\em action} of $G$ on $X$ is a homomorphism of $G$ into the symmetric group of $X$. To every element $x\in X$  and every element $g$ of $G$, 
there corresponds an element of $x\cdot g$ such that $(x\cdot g)\cdot h=x\cdot (gh)$ for any two 
elements $g$,$h$ of $G$, and $x\cdot e =x$, where $e$ is the identity element of $G$. 
	
An equivalence relation $\rho $ on $X$ is $G$-{\em invariant} or a {\em congruence}
on $X$ if the action of $G$ on $X$ preserves the relation, i.e., 
if $x=y(\mathrm{mod}\rho)$  then $x\cdot g =y\cdot g(\mathrm{mod}\rho)$
for all $g$ in $G$. 

A {\em $G$- set} is determined by the action of a group $G$ of  permutations of $X$, action that commutes with the projection map $\pi$. In other words, the action of $G$ is a fiber automorphism that permutes the fibers of $X$. \footnote{For G-sets, see \cite{Neumann}, {\em Chapter 3}, pp. 30-33. 
For congruences on $G$-sets, see ibid., {\em Chapter 7}, pp.74-75.}

If we take $G = Perm(S)$, the permutation group of $X$, there is a unique action of $Perm(S)$ on the total set $X$, action that commutes with $\pi$. The fibers of $X$ define a 
relation, of which they are the equivalence classes. So the fibered set $\pi$ is a $G$-set with a congruence and can be viewed the abstraction to sets with permutatins of a type of geometrical object.
\footnote{Important examples are principal bundles ( $G$ is a Lie group of diffeomorphisms) and regular coverings (or principal bundles with respect to a discrete group $G$ of transformations); see e.g., \cite{Dubrovin}}

\section{The Hole Argument for Sets}

With this machinery in place, it is easy to see that the hole argument that applies to 
fibered manifolds can be modified to apply to fibered sets. We can produce the fibered sets 
from the fibered manifolds by applying a forgetful functor \cite{MacLane1997} to the manifolds,  
that takes the category $\mathcal Mf_{n}$ of manifolds with diffeomorphisms in the category  $\mathcal Set$ of sets with  automorphisms.  Application of the 
forgetful  functor to the concept of a covariant theory results in the concept of a 
permutable theory. 

A  theory $\mathcal T$ is now a rule for selecting a class of sections of a fibered set $X$.  
The theory will be called {\em permutable} if, whenever a section $s:S \to X$  belongs to this 
class, so does  every section that results from applying a fiber automorphism to $s$. This 
results in a class of {\em automorphically-related} sections. The theory will be called 
{\em generally permutable} if all the  members of this class are {\em  semantically identical}.

As we have seen, the forgetful functor can be applied to the category of fibered manifolds in two 
ways: only to the base space, or to both the total space and the base space. we shall 
consider application of the forgetful functor each case in turn.

\subsection{Base Sets}

One can apply the forgetful functor to the base manifold only, abstracting from its 
topological and differentiable properties to 
get a base set $S \neq \emptyset$,  while allowing the fiber over each point $a \in S$ of the base set to remain a differentiable manifold $E_a$.

The total space is the product (union) of the fibers,
$\displaystyle E= \bigcup_{a\in S}E_a$. 

A cross section   
$\sigma : S \mapsto E$
takes each point $a$ on the base set $S$ into an element $\sigma (a)\in E_a$ of the fiber
over $a$
\footnote{While one could continue to talk of {\em points} of a fiber
in the applications to be discussued, each fiber will consist of a family of spaces, so it seems better to use the more neutral term {\em element} of a fiber.}.

At this level of abstraction, local diffeomorphisms\footnote{The distinction between diffeomorphisms and local diffeomorphisms disappears at this level of abstraction, 
as does that between global cross sections and local sections.}  
of the base manifold become 
permutations of the points of the base set, while fiber-preserving 
diffeomorphisms of the total space become fiber automorphisms, i.e., automorphisms of the fibers. 

As we shall now see, this case can be applied to the {\em quantum mechanics of many-particle systems}, and in 
particular can be applied to the cases in which these particles are all of the same kind.

Let the points of the base set be identified as elementary particles - 
for definiteness let us say a set of $N$ particles. The fiber over each such point 
represents a {\em state space} for that particle 
(later, we shall discuss various possible choices for such a state space). 
A {\em particular state} of the particle represented by an element of that state space. 
\footnote{"State of the particle" does not necessarily imply an instantaneous state. 
For example, in the Feynman approach be discussed below, an entire trajectory in 
configuration space represents a possible state of a particle.} 
The total space consists of the (ordered) product of each of the state-space fibers, 
and a cross section consists of a choice of state for each of the $N$ particles.
 Suppose all the particles are of the same natural kind (quiddity). 
Then the quantum statistics of elementary 
particles obliges us to postulate that, if one cross section is a possible state of the 
$N$-particle system, then any cross section that results from permutation of the fibers of the total space 
(or equivalently, from a permutation of the $N$ points of the base space) must also represent a 
possible state of the system. In our language, the theory must be permutable.

Now, we face the question of whether it must be generally permutable. 
In other words, in the face of quantum statistics, can we maintain the inherited 
individuality (hacceity) of particles of the same kind ? If we were to do so, then the hole argument would apply, 
and we would have to admit that any model of the theory must specify the individual state 
of each of the particles. But it is assumed in quantum mechanics that, in a case like this, 
one need only specify the overall state of the system, without any need to specify the state 
of each particle, considered as an individual. This, in effect, amounts to stipulating that the 
theory be {\em generally permutable} and that no individual characteristics are inherent in each
elementary particle, apart from those that it inherits from the overall state of the system.
\footnote{Auyang in \cite{Auyang}, pp. 162,  has emphasized the analogy 
between permutation of the labels of identical particles and coordinate transformations: 
Discussing  the permutation symmetry of the aggregate of particles , she says:
`` Conceptually, it is just a kind of coordinate transformation, where the coordinates are the 
particle indices. Permutation invariance is general and not confined to quantum mechanics. 
It says that specific particle labels have no physical significance.`` This is fine as 
far as it goes. But she does not discuss active permutations of the particles and the 
significance of {\em active} permutation invariance.}

What has been said so far could actually be applied to classical systems of identical particles. 
Quantum mechanics confronts us with a further complication, unique to it. 
There are states of the entire system that are not decomposable into products of $N$ one-particle states, even if we give up the individuality of thee 
particles. In fact, these {\em entangled states} are responsible for
the most characteristic features of quantum mechanics. 
Indeed, this problem arises,  independently of quantum statistics, even 
for a system of distinct elementary particles (i.e., each of a different kind) 
as seen in EPR type-correlation experiment.(see \cite{Stachel1997} for a discussion of these )

One way around this problem is to adopt the Feynman viewpoint, which allows us to maintain 
the concept of trajectories for each particle, 
but requires us to associate a {\em probability amplitude} with each trajectory and add the amplitudes 
for all trajectories that are not distinguishable within the given experimental 
setup.\footnote{See \cite{Stachel1997}, which contains references to Feynman's  papers} 
It is the addition of the amplitudes before ``squaring'' to find the probability of a process that is responsible for the entanglement of 
many-particle systems. To see this, consider the simplest possible many-particle system:
two particles not of the same kind. Then if the two are prepared in a certain way at a certain time, 
to calculate the probability that they will be detected in some way at a later time, we must consider all possible trajectories in the two-particle configuration space, consisting of the ordered product of two one-particle (ordinary) spaces.

Each one of these two-particle trajectories can be projected onto the two one-particle axes, 
so that it is associated with definite one-particle trajectories. But we must add the probability
 amplitudes for each of these two-particle trajectories in order to get an overall probability 
amplitude for the transition from the preparation to the registration.  And this overall 
probability amplitude is not associated with {\em any} one-particle amplitudes– rather, in some sense, 
it is associated with {\em all} of them.

To apply our technique to the Feynman approach, we must take as the total space the space of all 
possible paths in the  $N$-particle configuration space, which is the $N$-fold Cartesian
product of the $N$ one-particle configuration spaces. Each fiber is the space of all possible 
paths in the $(N-1)$-particle configuration space over one of the particles, an element of the 
fiber being one possible $(N-1)$-particle path. A section of the fibered space represents a possible 
path in the $N$-particle configuration space. All is well as long as the $N$ particles are of distinct kinds.
 But if they are all of the same kind, any theory applied to them must be generally permutable, 
and all of our previous results about such theories apply.
\footnote{If there is a Lagrangian for the system, it must be invariant under the
symmetric group $Sym(N)$, the action of each element of which is to permute the $N$ particles. 
If there is not a Lagrangian, the equations of motion must still be invariant under such permutations.}  
In particular, we could define the points of the base space (i.e., the particles) by starting from the total fibered space and identifying each fiber as a particle. 
Then, as in the manifold case, there are no independent permutations of the fibers and of the points of the base space, and the hole argument cannot even be formulated.

Another way to handle this problem is to introduce a reduced configuration space for identical 
particles by identifying all points in the $N$-fold Cartesian 
product of the $N$ one-particle configuration spaces that differ only by being permutations of the 
same $N$ points.\footnote{See \cite{Stachel1997}, section 7, pp. 252-253, which gives references 
to earlier work}  We can formulate this in terms of an equivalence relation on the original $N$-particle configuration space. Two points in this configuration space are equivalent if and only if one results from the other by a permutation of points of the $N$ one-particle configuration spaces. 
Then each equivalence class of points in the original configuration space corresponds to a point 
in the reduced configuration space.  

If and only if one excludes all collision points (i.e., points for which the coordinates 
of two or more of the particles are the same), this is a multiply-connected manifold, called the
reduced configuration space. It can be regarded as a quotient space of the $N$-particle configuration 
space by the symmetric group $Sym(N)$.
A trajectory in the reduced configuration space corresponds to a possible $N$-particle state that 
makes no distinction between the $N$ particles, so there is no need for 
(indeed no possibility of) further permutations.  In this, it is reminiscent of the 
Fock space treatment of particles in many-particle systems. As Teller in \cite{Teller}, pp. 51-52, 
emphasizes,  for bosons and fermions, respectively, ``A Fock space description ... provides
the most parsimonious description of multiquanta states\footnote{Teller prefers to 
use the words {\em quanta} and {\em multiquanta} instead of {\em particle} and 
{\em multi-particle} to emphasize the lack of inidividuality of the quantum objects.}...
this picture presents us with a view 
of entities quite free of primitive thisness.\footnote{I will use the word {\em thisness} 
for the property-transcending individuality that particles might be thought 
to have and that quanta do not have`` (\cite{Teller}, pp. 12). 
In the philosophical literature, this is often referred to as {\em haecceity}.
For further discussion in the context of this problem, see \cite{Stachel2005}.} ... it provides a description of the 
quantum mechanics of many-quanta systems of any kind, with no necessary tie to 
the field-theoretic setting or a relativistic description.``

\subsection{Fibered Sets}

If we apply the forgetful functor to the total space as well as the base space, 
abstracting from the topological and differentiable properties of both,
then the base space becomes a base set $S$, and the total space becomes a total set $X$, 
which is fibered by the base space $S$. A cross section of the fibered manifold becomes a 
section of $X$ over $S$. 
Local diffeomorphisms of the base manifold become automorphisms or 
permutations of the points of the base set, while fiber-preserving diffeomorphisms of the 
total space become fiber automorphisms of the total set. 
The resulting structure is very closely related to the concept of a congruence on a $G$-set
,and this relation is discussed in {\em Section 4}.
We shall forego discussing the hole argument for fibered sets in general, 
but turn to an important special case, relations between the elements of $S$.	

\section{The Relations Between Things}

\subsection{Sets, Sequences and Relations}

We shall now show that the concept of {\em fibered set} may be used to analyze relations 
between sets of various types of ``entities'' or ``things'' for short. 
We assume the set to be finite, although much of what is said would hold true even for sets of higher cardinality. 

Let {\bf S} be a set of $n$ entities, which we provisionally 
 label $\{a_{1},a_{2},..., a_{n}\}$ together with an ensemble 
$\mathbf{R}=\{R_{1},R_{2},..., R_{k}\}$ of $k$  
$n$-ary relations between these objects.\footnote{Characters 
in boldface will always stand for sets or ensembles (see next footnote) in Roman type for elements of a set, and in italic type for sequences of elements of a set} 

A pair ($\bf{S, R}$) is often called a {\em set with structure}.
\footnote{See \cite{Smith} pp. 3: Although it is usually used synonymously with {\em set}, we shall use the word {\em ensemble} to refer to a set-with-structure, i.e., one that may have some additional 
structure(s) relating its elements. ( `` Algebra can be characterized internally as the 
study of sets with structure `` - \cite{MacLane}, pp. 33). 
In particular, a set $S$ together with a multiset of relations {\bf R} on $S$ is called a 
{\em relational structure}. We shall leave open the question of the exact 
nature of this relational structure, not assuming a priori that the relations are 
either entirely independent of each other, so that their order is unimportant 
(as the word {\em set} would imply), nor that there is a unique order among them 
(as the word {\em sequence} would imply). They might, for example, 
have the structure of a partially ordered set, or something even more general.} 

Applying the hole argument to the category of fibered sets there are two possibilities:

1) The objects $\{a_{1},a_{2},..., a_{n}\}$  are individuated (i.e., distinguishable) 
without reference to the relations  $\mathbf{R}=\{R_{1},R_{2},..., R_{k}\}$.
\footnote{Note that this alternative leaves open the possibility that the distinction between these entities has been established by means of {\em some other} ensemble of relations (including properties, as one-place relations) 
between them, so long as this ensemble is entirely independent of {\bf R}}
In this case, the hole argument is valid.

2) The objects are {\em not} individuated (i.e. are indistinguishable, lack haecceity) without 
reference to the relations {\bf R}. In this case, the hole argument fails.

In order to formulate the hole argument for sets-with-
structure, we recall the mathematical definitions of  relation and sequence
(see e.g., \cite{Smith}).

Let $N$ be a natural number. For simplicity, in what follows we assume that $N\leq n$,
where  $n = \mathrm{card}S$, though it is possible to consider the case $N > n$.

An $N$-{\em sequence} of elements of a nonempty set $S$ is 
defined as a function  $s:\{1,2,...N\}\to S$. The elements of the sequence are denoted by 
$s(i)=s_i$ for $i\in \{1,2,...N\}$. Duplications of elements is allowed
i.e., $s_i = s_j$, where $i,j\in \{1,2,...N\}$.    

The direct (Cartesian) product set $S^{N}=S\times S\times ...\times S$, ( $S$ is taken $N$ times). 
The elements of $S^N$ are isomorphic to $N$-uples of elements from the set $S$.

An $N$-ary relation( or often simply a relation) is a generalization of binary relations such as 
``$=$'' and ``$< $'' which occur in statements such as "$2 < 3$" or "$1 + 1 = 2$". 
or $R$ is 3-ary  relation on $Z\times Z\times Z$ consisting of 
$(a_{1}, a_{2}, a_{3})$ such that $a_{i}\in Z$ and $a_{1}<a_{2}<a_{3}$. 
It is the fundamental notion in the relational model for databases. 

An $N$-ary relation
$R$ is defined as a subset of $S^N$ i.e.,
$$R=\{(a_{1},a_{2},...,a_{N})\}\subseteq S^N.$$ 

An $n$-ary predicate is a {\em truth-valued} function of $n$ variables. 

Because a $n$-ary relation defines uniquely an $n$-ary predicate that holds for $a_{1}, ..., a_{n}$ if $(a_{1}, ..., a_{n})$ is in $R$, and vice versa, the relation and the predicate are often denoted with the same symbol. 

The relation $R$ holds between the sequence of objects $(a_{1},a_{2},..., a_{n})$ 
will be abbreviated  ''$R(a_{1},a_{2},..., a_{N})$ holds or simply "$R(a_{1},a_{2},..., a_{N})$" or 
``$(a_{1},a_{2},..., a_{n}\in R$"

If the (ordered) sequence $(a_{1},a_{2},..., a_{N})$ does not belong to this subset $R$, 
the proposition is false.  We shall abreviate this by ''$R(a_{1},a_{2},..., a_{N})$ doesn't hold``or $(a_{1},a_{2},..., a_{n})\not\in R$''

Note that any $N$-ary relation with $N < n$ can be treated as an $n$-ary relation by 
simply adjoining to the subset of $S^N$ defining the desired relation {\em all} subsets 
of  $S^{n-N}$. 
Then, only the entities occupying of the first $N$ places in a sequence of $n$ entities will determine whether the $N$-ary relation holds. So from now on, we shall only consider $n$-ary relations, i.e., subsets of $S^n$, with the understanding that this constitutes no loss of generality.

A {\em totally symmetric} $n$-ary relation $S^{n}$ is a $n$-ary relation $R\subseteq S^{n}$ such that
if $R(a_{1},a_{2},..a_{n}) $ holds, then $R(a_{i{1}},a_{i_{2}},..a_{i{n}}) $ holds for all permutations
$(i_{1},i_{2},..i_{n})$ of the indices $(1,2,...,n).$ \footnote{In \cite{SSaunders} Saunders shows that if the $n$ entities are indistiguishable, then nothing is lost by replacing a $n$-ary relation wih a totally symmetric $n$-ary relation}

Below are some other examples of relations that we use in the paper:

An {\em equivalence relation} is a binary relation that is reflexive, symmetric and transitive (RST). It induces a partition of the set into equivalence 
classes of subsets that are mutually exclusive and exhaust the set. 

A {\em partial ordering} is a binary relation that is reflexive, antisymmetric and transitive (RAT).
A poset  (partially ordered set) is one example of a {\em set-with-structure}, and all approaches to space-time theory based on posets, in particular Sorkin's causal set approach 
\cite{Sorkin}, \cite{K-Penrose} fall within the purview of this treatment of sets-with-structure. 
In particular, the hole argument developed below applies to all such approaches.

\subsection{Fibered Sets and Relations}

We are ready to define the type of fibered set, the cross-sections of which correspond to relations.
Consider the base set  $S$, and the total set $E=S^n$. 
The projection is defined as follows: $\pi: E \to S$, 
the fiber over some element $a\in S$ is the set of all $n$-uples $(a, a_{i_{2}}, ..., a_{i_{n}})$ of elements of $S$ that begin with that element $a$. That is, the $n$-uple 
$(a_{1}, a_{2}, ..., a_{n})$ belongs to the fiber over $a$ if and only if $a_{1} = a$. 
The projection operation is thus the projection over the first set, $\pi=pr_1$ where
$pr_{1}(a_{1},a_{2},...a_{n})=a_{1}$.
The {\em fiber over} $a$, $\pi ^{-1}(a)=\{(a,a_{2},...a_{n})/ a, a_{i}\in S\}=\{a\}\times S^{n-1}$ 
is isomorphic to $S^{n-1}$

A {\em cross-section} of $\pi$ is a map $\sigma :S\to S^{n}$ such that $\pi\circ \sigma =\mathrm{id}_{S}$.
A cross-section $\sigma $ can also be interpreted as the graph of a function from $S$ to the fiber $S^{n-1}$.

We have a natural fiber preserving automorphism $f:S^{n}\to S^{n}$ associated to every automorphism $f_S$ of the base 
set $S$. 
Namely, if $f_{S}(a)=b$ under a base automorphism $f_{S}:S\to S$, then
$f(a,a_{i_{2}},...a_{i_{n}})=(b, b_{i_{2}},...b_{i_{n}})$, where  $b_{i}=f(a_{i})$ for all $i$.

These fiber automorphisms provide a representation of the permutation group acting on the base 
set.  Clearly, if  $f_{S}(a)=b$ and $g_{S}(b)=c$  under base automorphisms, 
the corresponding fiber automorphisms then satisfy
$f(a,a_{i_{2}},...a_{i_{n}})=(b, b_{i_{2}},...b_{i_{n}})$ and $g(b, b_{i_{2}},...b_{i_{n}})=
(c, c_{i_{2}},...c_{i_{n}})$, 
so that $(g\circ f)_{S}(a)=c$ corresponds to the fiber automorphisms that takes
$(g\circ f)(a,a_{i_{2}},...a_{i_{n}})=(c, c_{i_{2}},...c_{i_{n}})$. A similar argument holds for inverse automorphisms. To the   identity base automorphism $id_{S}(a)=a$ corresponds the identity fiber automorphism $id: S^{n}\to S^n$, $id(a,a_{i_{2}},...a_{i_{n}})=(a,a_{i_{2}},...a_{i_{n}})$, because $id(a_{i})=a_i$  for all indices $i$.

We assert the following:

\vspace{5pt}
{\bf Proposition 5.1}\\ 
{\em A cross-section of $\pi: S^{n}\to S$  defines an $n$-ary relation over $S$.}\\
\vspace{5pt}
The proof is a consequence of the definitions of an $n$-ary relation and  $\pi$.\\

If we denote with  $E'$ the reduced fibered set $(S\times [S^{n-1}/Sym(S)]$, and 
where the fiber $S^{n-1}/Sym(S)$ is the quotient space
\footnote{see pp. 23} of $S^{n-1}$ by the symmetric group of the set $S$. 
Then entire equivalence class of cross-sections of the fibered set $(\pi: S^{n} \to S)$
corresponds to one cross section of the reduced fiber set 
$(E'\stackrel{pr_{1}}\to S)$.

We have the following proposition.

\vspace{5pt}
{\bf Proposition 5.2}\\
{\em An $n$-ary symmetric relation $R$ is a cross-section of the reduced fibered set \\
$(E'\stackrel{pr_{1}}\to S)$.}

{\bf Proof}  

Let $X\subseteq S^{n}$ be an $n$-ary symmetric  relation on $S$. 
An element of $X$  is a sequence of elements of the base space, begining with some element 
of the base space, let us say ``$a$''. All the elements of $X$ consisting of sequences that begin with the same element $a$ of the base space form a subset $X_{a}$ of $X$.
By the definition given above of a fibered set, $X_{a}$  lies on the fiber over $a$ of that set.
$X$ is the union of all the subsets  $X_{a}$ of $X$, each of which lies on some (different) fiber of the fibered set. 
But there may be fibers of the set that do not have any element of $X_{a}$ on them. In that case, we choose the null set $\emptyset$ on that fiber. Now every fiber has one (and only one) element chosen on it, so that the chosen elements constitute a class of cross-sections of  the reduced fibered set induced by a fibre automorphism of $S^n$ that projects on the identity map $\mathrm{id}_{S}:S\to S$.
Now as mentioned above, an entire equivalence class of cross-sections of the
fibered set corresponds to one cross section of the reduced fiber set.  We have thus proved that
an $n$-ary symmetric relation $X$ is a cross-section of the reduced fibered set $(E'\stackrel{pr_{1}}\to S)$.

\subsection{The Hole Argument for Relations}

It is easy to see now how the original hole argument can be formulated at the level of sets with structures, and permutations of the set members and their relation.
A theory is a rule that selects a class of worlds, a class of ensembles of $n$-ary relations
$\mathbf{R}=\{R_{1},R_{2},..., R_{k}\}$ whose places are filed by the members of the same set $S$
 of $n$ entities {\bf a}. Further we will consider a permutable theory, i.e. if {\bf R} is in the selected class of worlds, so is $P\mathbf{R}$ for all $P$.
If $\mathbf{R(a)}$ represents a possible state of the world, could such a permutable theory select a 
{\em unique} state of the world, by first specifying a unique world (i.e. one {\bf R} ) and then specifying any number $N$ of its places less than $n-1$ are filled?

The theory of {\em possible worlds} originated in Leibniz' ideas, and has been developed in recent times by analytical philosophers such as Kripke \cite{Kripke} and Lewis \cite{Lewis} as a method of
solving problems in formal semantics.  The concept of possible worlds is related to modal logics ( see e.g. \cite{Popkorn}) and is used to express modal claims. The two basic modal operators are $\square$ (necessity) and  $\diamond$ (possibility).  
The mathematics of the theory of possible worlds is {\em set-theory and relations}.  
\footnote{In a  Kripke model., there is a central element interpreted as 'the actual world'. A world is possible if it is related to the central element by  an 'accessibility relation', which is a binary relation. In general, the distinction between possible and impossible worlds depends on the definition of the accessibility relation. The most disputed problem is the nature of the property that defines one world as actual.\cite{Lewis}}

We call a relational structure $(S, \mathbf{R)}$ a world. There are $n!$ permutations of  $n$ entities forming the permutation group $Sym(S)$, 
so there are $n !$ possible sequences of the members of $S$. 
Let $\mathbf{a}=(a_{1}, a_{2},...a_{n})\in S^{n}$  be one such sequence. We write $P\mathbf{a}$ 
to symbolize a permutation of the sequence $\mathbf{a}$, and $\mathbf{Pa}$ to symbolize the entire set of them.

We write $R(\mathbf{a})$ as an abbreviation for the $n$-ary relation $R$ with its places filled by some 
definite sequence  {\bf{a}} of the $n$ entities, and $\mathbf{R(a)}$
for the entire ensemble of  relations with their places filled in that sequence. 

We see that for some $\mathbf{a}=(a_{1}, a_{2},...a_{n})\in S^{n}$ , the ensemble of $n$-ary relations 
$\mathbf{a}\in \mathbf{R}$, while for other $n$-tuples, $\mathbf{a}\not\in \mathbf{R}$.
\footnote{This means that the assertion Ri(a) is either true or false, 
but not meaningless. Note that our stipulation leaves open the possibility that the distinction between these entities is established by 
means of some other ensemble of relations between them, which is entirely independent of  {\bf R}.} 
We call $\mathbf{R(a)}$ a {\em possible state of the world}, whether or not $\mathbf{R(a)}$.
If $\mathbf{a}\in \mathbf{R}$ for any sequence $\mathbf{a}$ in $S^n$, then $\mathbf{R(a)}$ is called a {\em state of the world}.

Consider  an ensemble of $n$-ary relations $\mathbf{R}=(R_{1}, R_{2},...R_{k})$ on $S$. 
Let $\mathbf{a}=(a_{1}, a_{2},...a_{n})\in S^{n}$ and $P:S \to S$ a permutation on $S$. Then $Pa$ is a 
permutated sequence of the $n$-uple $\mathbf a$. Then if $\mathbf{R(a)}$ is a  possible state of the 
world then $\mathbf{R(Pa)}$ is another possible states of the world. 
But, in general, if $\mathbf{a}\in \mathbf{R}$ then does not necessary imply 
that $\mathbf{a}\in \mathbf{R}$. 
Therefore we define another ensemble of relations, denoted $\mathbf{PR}=(PR_{1}, PR_{2},...PR_{k})$
which holds for $a$ if and only if $\mathbf{R}(P^{-1}a)$, where  $P^{-1}$ is the 
permutation inverse to $P$.
That is, the two ensembles of relations are elementarily equivalent, which we write  
$\mathbf{PR}(\mathbf{a})=\mathbf{R}(P^{-1}\mathbf{a})$ or $PR_{i}(\mathbf{a})=R_{i}(P^{-1}\mathbf{a})$ for all $i\in \{1,2...n\}$

It follows trivially from this definition (by substituting $P\mathbf{a}$ for $\mathbf a$ in this equation,
and noting that $P\circ P^{-1}= Id$ , the identity permutation), that $P\mathbf{R}(P\mathbf{a})$ 
if and only if $\mathbf{R(a)}$ – that is, they are too elementarily equivalent, 
i.e.,  
$P\mathbf{R}(Pa)=\mathbf{R(a)}$ or $PR_{i}(P\mathbf{a})=R_{i}(\mathbf{a})$ holds for all $i\in \{1,2...n\}$.

This trivial identity – or,  if you will, tautology (since it depends only on the definition of 
$\mathbf{PR}$) has been taken to contain the essence of  covariance and hence to 
have important bearing on the 
hole argument; \footnote{See \cite{Earman1997}.  
Of course, their discussion refers to the diffeomorphism version of this identity, discussed in Section 1. But the permutation version clearly captures the essence of the matter at a much higher level of abstraction: The point is that a diffeomorphism is just a 
highfalutin version of a permutation, as we shall discuss further below.} 

In fibered sets formulation, this trivial identity can be interpreted as follows: we have a natural fiber preserving automorphism, denoted
$\hat{P}:E'\to E'$ of the reduced fiber set $\pi'=pr_{1}:E'\to S$. 
We see that  $\hat P$  takes
a section  $\sigma:S\to E'$ ( i.e, a $n$-ary symmetric relation $\mathbf{R(a)}$ on $S$) 
to the carried-along section 
$\bar\sigma:S\to E$, $\bar\sigma=\pi'P\circ\sigma\circ P^{-1}$ (i.e. the $n$-ary symmetric relation
$P\mathbf{R}(P^{-1}(\mathbf{a}))$ ).

Therefore, there is a natural homomorphism\footnote{The natural homomorphisms in Kripke semantics are called $p$-morphisms defined as a map between worlds that preserves the accessibility relations.} 
from $(S,\mathbf{R})$ to $(S,\mathbf{PR})$ which is defined to be a bijective mapping $P:S\to S$ such that $P$ preserves the relations, i.e.'' $\mathbf{R(a)}$'' implies ''$\mathbf{PR(a)}$''.
In other words, $(S,\mathbf{R})$ is a possible world if and only if $(S,\mathbf{PR})$ is a possible world.

A world {\bf (a ,R)}is called {\em permutable} if the following condition holds: 
If {\bf R(a)} is a {\em possible state of the world},  then {\bf PR(a)} is also a possible state of the 
world, for all permutations $P$ of {\bf a}. This clearly sets up an equivalence relation between possible states of the world: two possible states of the world are equivalent if they differ by a 
permutation. It will be {\em generally permutable}, if all of these possible states are elementarily equivalent.

\section{Conclusion}

As we have seen in {\em Section 3}, in general-relativistic theories, 
starting from some natural bundle ,
the points of space-time may be characterized as such independently of the particular relations in which they stand; but they are entirely individuated in terms of the relational structure given by 
cross-sections of some fibered manifold. And we have seen in {\em Section 5} that  the elementary particles are similarly individuated by their position in a relational structure. Electrons as a kind may be characterized in a way that is independent of the relational structure in which they are imbricated by their mass, spin and charge, for example; but a particular electron can only be individuated by its role in such a structure. The reason for this claim is, of course, the requirement that all relations between N of these particles be invariant under the permutation group acting on these particles.  As we have seen, it is  possible to make the same move in both the case of the fibered manifolds that generalized space-time and the fibered sets used to describe elementary particles: by defining the elements of the base set as corresponding to the equivalenc fibers of the fiber space, we make it impossible to carry out a fiber automorphism (i.e, to permute these fibers) without carrying out the corresponding automorphism (i.e., permuting the elements) of the base space.			
This suggest the following viewpoint: Since the basic building blocks of any model of the universe, the elementary particles and the points of space-time, are individuated entirely in terms of the relational structures in which they are embedded, only "higher-level" entities constructed from these building blocks can be individuated independently. Therefore, the following principle of generalized covariance should be a requirement on any fundamental theory: The theory should be invariant under all permutations of the basic elements, out of which the theory is constructed.

Perturbative string theory fails this test, since the background space-time (of no matter how many dimensions) is only invariant under a finite-parameter Lie subgroup of the group of all possible diffeomorphisms of its elements.

This point now seems to be widely acknowledged in the string community. I quote from two recent review articles. Speaking of the original string theory Michael Green\cite{Green} notes: 
 
``This description of string theory is wedded to a semiclassical perturbative formulation in which the string is viewed as a particle moving through a fixed background geometry .... Although the series of superstring diagrams has an elegant description in terms of two-dimensional surfaces embedded in spacetime, this is only the perturbative approximation to some underlying structure that must include a description of the quantum geometry of the target space as well as the strings propagating through it ( p. A78). ... A conceptually complete theory of quantum gravity cannot be based on a background dependent perturbation theory .... In ... a complete formulation the notion of string-like particles would arise only as an approximation, as would the whole notion of classical spacetime   (p. A 86) ''

Speaking of the more recent development of M-theory, Green says:

``An even worse problem with the present formulation of the matrix model is that the formalism is manifestly background dependent. This may be adequate for understanding M theory in specific backgrounds but is obviously not the fundamental way of describing quantum gravity (p. A 96).''

And in a review of matrix theory, {\em Thomas Banks} comments: 
 
String theorists have long fantasized about a beautiful new physical principle which will replace Einstein's marriage of Riemannian geometry and gravitation. Matrix theory most emphatically does not provide us with such a principle. Gravity and geometry emerge in a rather awkward fashion, if at all. Surely this is the major defect of the current formulation, and we need to make a further conceptual step in order to overcome it (pp. 181-182).
 It is my hope that emphasis on the importance of the principle of dynamic individuation of the fundamental entities, with its corollary requirement of invariance of the theory under the entire permutation group acting on these entities, constitutes a small contribution to the taking of that further conceptual step.

\section{Acknowledgements} This paper has been substantially improved in response to the oral comments of Abhay Ashtekar at the  presentation of an earlier version at Pennsylvania State University by John Stachel, who thanks Abhay for his hospitality as well; and to written comments by the late Armand Borel on an earlier draft and especially to the extensive written critique of that draft by Erhard Scholz. 

\section{Appendix}

In {\em Section 2}, it was simply postulated that, to every diffeomorphism of the base space 
there corresponds a unique fiber automorphism, and cross sections of fibered manifolds with 
this property were used to define geometric objects. 
It is worthwhile to see in detail how such a fiber automorphism may be constructed by lifting a 
local diffeomorphism of $M$ into a tensor bundle automorphism, 
and thereby connect up with the usual definition of a tensor. 
We do the lifting with the help of a linear frame field in $M$. 

A linear frame $f$ in an $n$-dimensional vector space is an ordered set of linearly independent vectors.

A linear frame  $f_{p}=(e_{i})_{i=1,..n}$ at a point $p$ of an $n$-dimensional manifold $M$ is a basis for the tangent space $T_{x}M$. 
The collection of all linear frames at all points of $M$ forms the total space of a fibre bundle 
$LM$, called the frame bundle.  Its base is $M$, the projection is the map that associates to each 
frame the point in which the tangent space lies and its standard fibre is the linear group
$GL(n, \mathbf{R})$.

A local linear frame field $f$ is merely a local section of the frame bundle $LM$. 
In other words, a frame field on some open set $U\subseteq  M$ are $n$-vector fields 
$f=(e_{i})_{i=1,..n}$ on $U$, such that $f(p)=(e_{i}(p))_{i=1,..n}$  is a linear frame at 
each $p$ in $U$.

Suppose the points of $M$ are subject to some diffeomorphism $D:M\to M$, $D:p\longmapsto Dp$.

In physics only the physical components of a tensor ( or more generally, of a geometric object) 
can be measured, i.e. the {\em scalars} gotten from projection of the tensor onto a suitable choosen tangent basis and its dual cotangent basis, only these can be measured \cite{Pirani} .

The values of a scalar field $\lambda $ at two different points $p$ and $Dp$ can be compared without 
further ado since they are just numbers: $\lambda [p]$ and $\lambda [Dp]$.

If $F[p]$ is not a scalar field, we must take its components $F_{f[p]}$ with respect to some 
frame $f[p]$ at point $p$ in order to get sets of numbers characterizing the geometrical 
object field that we can take from one point to another of $M$ for comparison. 

Given a frame $f_p$, the diffeomorphism $D$ induces a corresponding new frame $D^{*}f_{Dp}$ at the point $Dp$. This is called the frame {\em carried (or dragged) along} with the diffeomorphism $D$.

Let $F$ be a field. The new field $D^{*}F$, called the {\em pullback} of $F$, is defined as follows:
$D^{*}F_{f}[p]:= F_{D^{-1}f}[D^{-1}p]$.
In other words: the components of the pullback field with respect to the frame at a point 
are numerically equal to the components of the original field in the pulled-back frame at 
the pulled-back point. 

It is shown that this definition is independent of the particular linear frame 
originally used to define the components of the geometrical fields.

Interchanging the roles of $D$ and $D^{-1}$ in the above expression we obtain 
an equivalent expression $(D^{-1})^{*}F_{f}[p]= F_{Df}[Dp]$.
In other words, whether we carry out a diffeomorphism without carrying along the $F$ 
(right hand side), or pull back the $F$ by the inverse transformation without carrying out a 
diffeomorphism (left hand side), the result is the same \footnote{``Don't raise the bridge, lower the 
river !''}.

By substituting now $Dp$ for $p$ in the definition of $D^{*}F$, the above formula can be put in 
another equivalent form: $D^{*}F_{Df}[p]= F_{f}[p]$.
So, the values of the carried-along components with respect to the
carried-along frame at the carried-along point are numerically equal to the values of 
the original field with respect to the original frame at the original point. 

Since, as noted above, any diffeomorphism $D$ induces a corresponding change of basis field, 
it follows that the model $(M, D^{*}F[Dp])$ is equivalent to the original model $(M, F[p])$.
(``If we drag everything, we change nothing.``) 

Since it follows from the definition of $D^{*}F$, this equivalence cannot 
fail to hold; so I shall refer to it as the trivial equivalence.
Now I shall indicate how the hole argument runs in this context. 

Suppose we impose the following requirement on a theory $\mathcal T$ :
For all diffeomorphisms $D$,  if $(M, F[p])$ is a model, then so is  $(M, D^{*}F[p])$
(note that we do {\em not} require the two models to be equivalent). 
In particular, this will be the case if the models are defined as the solutions to a set of 
generally-covariant field equations for the F[p] fields. 
\footnote{Indeed, we may take this as the definition of 
such a set of generally-covariant field equations.} We shall call a theory {\em covariant} if 
it obeys this requirement. Remember that, by the trivial identity,  $(M, F[p])$ and $(M, D^{*}F[Dp])$
always represent elementarily equivalent models, and thus do so for a covariant theory quite independently of the hole argument for such theories, which I now proceed to discuss. \footnote{As discussed earlier, 
this view is in sharp contrast with that expoused in \cite{Earman1987}, who 
also denote as ``generally covariant`` theories that we call ``covariant.`` }

The hole argument hinges on the answer to the following question. In a covariant theory, 
is it possible to pick out a {\em unique} model by specifying $F[p]$ everywhere except on some open  
submanifold $H$ of $M$ (the "hole"),  i.e., on $M -H$? If  $(M, F[p])$ and $(M, D^{*}F[p])$
 are {\em inequivalent} models for all $D$ except the identity diffeomorphism (by the definition of a 
covariant theory, {\em all} of them are models of the theory if one is), then the answer is no. 
For we can then pick any diffeomorphism $D_H$ that is equal to the identity diffeomorphism on
 $M -H$, but differs from the identity diffeomorphism on $H$. 
Then $(M, F[p])$ and  ($(M, D^{*}F[p])$ will be two {\em different} models that agree on $M-H$ 
but differ on $H$, and are therefore inequivalent models. So no conditions imposed on $F[p]$ on 
$M -H$ can serve to fix a unique model on $H$.

\end{document}